# Single-molecule Surface-Induced Fluorescence Attenuation Based on Reduced Graphene Oxide*

Qin-Kai Fan [1,2]　　Chen-Guang Yang [1,2]　　Shu-Xin Hu [1]　　Chun-Hua Xu [1]
Ming Li [1,2]　　Ying Lu [1,2] †

[1] Institute of Physics, Chinese Academy of Sciences, Beijing National Laboratory for Condensed Matter Physics, Key Laboratory of Soft Matter Physics, Beijing 100190, China
[2] University of Chinese Academy of Sciences, Beijing 100049, China

## Abstract

Single-molecule surface-induced fluorescence attenuation (smSIFA) is a precise method for studying the vertical movement of biological macromolecules using two-dimensional material acceptors. Unlike other methods, smSIFA is not influenced by the planar motion of membranes or proteins. However, the detection range and accuracy of vertical movement are dependent on the properties of these two-dimensional materials. Recently, smSIFA utilizing graphene oxide and graphene has significantly advanced the study of biomacromolecules, although the detection range is restricted by their inherent quenching distances. Modifying these distances necessitates the replacement of the medium material, which presents challenges in material selection and preparation. Consequently, there is a pressing need to develop controllable materials for smSIFA applications. In this study, we enhance the smSIFA technique using graphene oxide as the medium acceptor through thermal reduction. By adjusting the reduction temperature, we prepare reduced graphene oxides at varying degrees of reduction, thus fine-tuning the quenching distances. The adjustment of these distances is measured using fluorescently labeled DNA. This modified smSIFA approach, employing reduced graphene oxide, is then applied to observe conformational changes in the Holliday junction, demonstrating the enhanced detection capabilities of reduced graphene oxide.

## 1. Introduction

Single-molecule fluorescence resonance energy transfer (smFRET) is a commonly used method for studying the dynamic processes of biomacromolecules, offering high temporal and spatial resolution[1–4]. Membrane proteins are the main functional components of biological membranes, accounting for approximately one-third of the genes in a cell[5]. Studying the orientation and insertion depth of membrane proteins in the cell membrane is crucial for understanding their structure and function[6,7]. Due to the fluidity of the cell membrane, membrane proteins on the cell membrane undergo lateral movement, limiting the application of FRET in the study of membrane proteins[8−10].

In recent years, surface-induced fluorescence attenuation (SIFA) technology, based on the principle of fluorescence resonance energy transfer, has been developed. This technology utilizes two-dimensional materials as fluorescence acceptors. By analyzing the degree of fluorescence intensity attenuation of the fluorescence donor, it can calculate the vertical distance between the donor and the material surface[11,12]. The dependency between the fluorescence attenuation efficiency and the vertical distance between the donor and the material surface is described by Equation (1):

$$E_{\text{SIFA}} = 1 - I/I_0 = 1/(1 + (d/d_0)^4) , \qquad (1)$$

where $I_0$ is the fluorescence intensity of fluorescence donor in the absence of material; $I$ the intensity of fluorescence donor in the presence of material; $d_0$ is the characteristic quenching distance at which $I/I_0$=0.5. When selecting two-dimensional materials as fluorescence acceptors, it is necessary to consider not only their biocompatibility but also their optical properties and ease of preparation. In recent years, SIFA based on graphene oxide (GO) has been widely used in the study of membrane proteins[13,14], and significant progress has been made in single-molecule fluorescence imaging based on graphene[15,16]. Previous studies have shown that $d_0$ of GO is 4 nm[11], and $d_0$ of graphene is approximately 18 nm[16−18]. GO has a smaller $d_0$, suitable for detecting signals within a range of 2−6 nm from its surface[11]; graphene has a larger $d_0$, suitable for detecting signals beyond 12 nm from its surface[16]. Figure 1(a) illustrates the limitations of using GO and graphene as medium acceptors, demonstrating that they cannot completely cover the area near the cell membrane surface, leaving undetectable regions. Figure 1(b) shows the distance quenching curves for different $d_0$ values. The slope of the attenuation curve is the highest near $d_0$, providing the highest detection sensitivity. The detection sensitivity decreases significantly when the fluorescence donor is too close or too far from the surface. The $d_0$ of different two-dimensional materials varies, determining the detectable range. When performing SIFA experiments, it is necessary to select a material with the target $d_0$ based on the size of the target macromolecule and the position of the fluorescent labeling site.

The detection of the vertical distance changes of the fluorescence donor from the cell membrane surface in SIFA methods is achieved by measuring the intensity of the donor. Therefore, the sensitivity of SIFA is limited by the instrument's ability to detect the intensity of the fluorescence donor. It is generally believed that the intensity change ratio that the instrument can resolve is approximately 10%[11]. Based on this, the distance detection sensitivity of SIFA for different $d_0$ values can be calculated using Equation (1). Figure 1(c) shows the distance detection sensitivity curve of SIFA. GO has $d_0$ = 4 nm, and the detection sensitivity near $d_0$ can reach approximately 0.5 nm. Within a range of 2−6 nm from its surface, GO can achieve a resolution of 1 nm. However, beyond 3−6 nm, the detection resolution of GO drops to 3 nm and decreases exponentially with increasing distance. When $d_0$ increases to 10 nm, although it can detect fluorescence signals from a farther range from the surface, even at the most sensitive $d_0$, it can only achieve a detection precision of 1 nm. When performing SIFA experiments, if a spatial resolution of 1 nm is to be achieved, the $d_0$ of the medium acceptor should not exceed 10 nm. As shown in Figure 1(c), to detect biomacromolecules within a range of 12 nm from the surface of the medium and compensate for the limitations of graphene and GO, while ensuring a spatial resolution of 1 nm, it is necessary to continuously adjust $d_0$ within the range of 4−10 nm. Preparing surface materials with $d_0$ values between 3−10 nm and applying them to the observation of biomacromolecules in SIFA experiments poses challenges in material selection and preparation. Studies have shown that oxidizing graphene by applying an electric current can reduce $d_0$, but the oxidation of graphene is uneven[19]. Reduced graphene oxide (rGO) is obtained by reducing GO through heating, chemical, and electrical methods. rGO is easy to prepare and has physical and chemical properties similar to graphene[20−22]. Studies have shown that the fluorescence quenching effect of rGO is between that of GO and graphene[23,24], but rGO lacks applications in single-molecule fluorescence imaging. In this study, we prepared rGO using thermal reduction and applied it to single-molecule fluorescence imaging. We precisely measured the $d_0$ of rGO with different reduction degrees using fluorescently labeled double-stranded DNA rulers, improved the existing SIFA technology, and verified the advantages of rGO-SIFA technology by comparing it with GO.

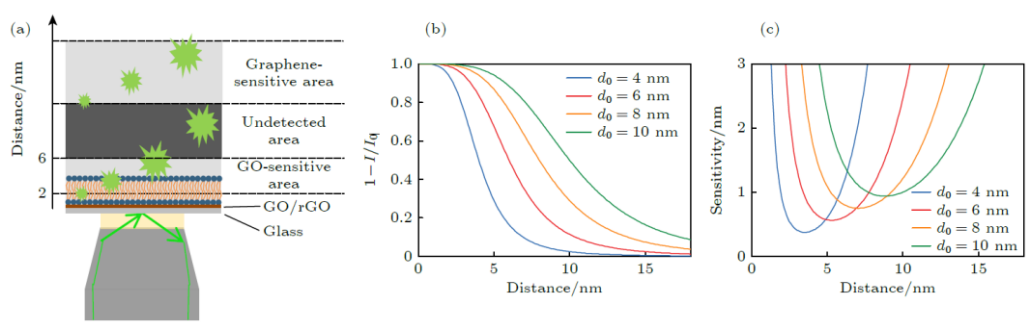

Fig.1. SIFA method of regulatable $d_0$. (a) Schematic representation of SIFA; (b) The relationship between degree of attenuation of a fluorescent donor and donor-surface distance; (c) The relationship between detection sensitivity of SIFA and donor-surface distance.

## 2. Experimental Materials and Methods

2.1 Preparation of GO Dispersion
The preparation of GO dispersion was performed according to the Hummers method [25,26]. 1 g $NaNO_3$, 46 mL $H_2SO_4$, and 1 g graphite were mixed uniformly at 0°C and kept in an ice bath to maintain a temperature of 0°C. 6 g $KMnO_4$ was slowly added to the mixture with continuous stirring for about 2 h. Then, the mixture was transferred to a 35°C environment for continuous stirring for 2 h. Subsequently, 120 mL deionized water was slowly added to the mixture (over 30 min), followed by the slow addition of 6 mL $H_2O_2$ (30% by mass) and incubation for about 20 min. The mixture was centrifuged at high speed (15000 r/min) for 20 min, and the supernatant was removed. The precipitate was then redispersed in deionized water and subjected to ultrasonic treatment (100 W, 30 min) to exfoliate the multilayer GO. After low-speed centrifugation (1000 r/min) for 10 min, the precipitate was removed, and the supernatant was retained. This step was repeated 3−4 times until no obvious precipitate appeared. The collected solution was further centrifuged, and the supernatant was removed after each centrifugation, and the precipitate was redispersed for the next centrifugation. After centrifugation at 8000, 6000, and 4000 r/min for 25 min each, the precipitate was redispersed and centrifuged again at 2000 r/min for 25 min. The supernatant obtained was the GO dispersion used in the experiment.

## 2.2 XPS Analysis of GO Reduction

The GO dispersion was vacuum-filtered using a nitrocellulose membrane (diameter: 47 mm, pore size: 0.2 μm, Whatman) to obtain a GO film [27]. The film was naturally dried and then separated from the filter membrane. A small piece of GO film was cut and sandwiched between two clean coverslips, which were placed in a vacuum tube furnace for baking. The heating rate was set to 8°C/min, and the temperature was maintained at the target temperature for 2 h before naturally cooling to room temperature. The rGO film after heating reduction and the GO film were analyzed using X-ray photoelectron spectroscopy (XPS, Thermo Fisher Scientific ESCALAB 250X) at the Institute of Physics, Chinese Academy of Sciences. The X-ray source used was monochromatic Al Ka X-rays with an energy of 1486.6 eV.

## 2.3 rGO-SIFA Experiment

The rGO-SIFA experiment consists of three parts: preparation of the rGO-SIFA sample chamber, preparation of single-molecule fluorescence experimental materials and samples, and fluorescence imaging.

### 2.3.1 Preparation of the rGO-SIFA Sample Chamber

The cover glass used in the experiment needed to be cleaned to remove surface fluorescence impurities that could interfere with the single-molecule fluorescence imaging experiment. The cleaning process involved ultrasonic cleaning with acetone for 30 min, followed by ultrasonic cleaning with deionized water to remove acetone residue, then ultrasonic cleaning with methanol for 30 min, and finally ultrasonic cleaning with deionized water to remove methanol residue after ultrasonic cleaning. The cover glass was ultrasonically cleaned with 1 mol/L NaOH solution for 10 min each time, and the NaOH solution was replaced after each cleaning. This process was repeated 3 times to obtain clean cover glass without fluorescence impurities.

Using the Langmuir-Blodgett (LB) technique, single-layer GO from the GO dispersion was transferred onto the cleaned cover glass[28]. Another clean cover glass was placed on top, and the assembly was placed in a vacuum tube furnace for baking. The heating rate was set to 8°C/min, and the temperature was maintained at the target temperature for 2 h before naturally cooling to room temperature. The rGO cover glass was then attached to a clean slide using double-sided tape to create a sample chamber suitable for loading on a total internal reflection fluorescence microscope.

### 2.3.2 Preparation of Single-Molecule Fluorescence Experimental Materials and Samples

The DNA sequences used in the experiment were purchased from Sangon Biotech (Shanghai) Co., Ltd. The sequence of the double-stranded DNA was TATGGTCAACTGCTGAGCGTAG-biotin, ATACCAGTTGACGACTCGCAT. The sequences of the four DNA strands that make up the DNA Holliday junction are shown in Table 2. The annealing buffer (pH = 7.5) contained 50 mmol/L NaCl and 25 mmol/L Tris-HCl. For annealing the double-stranded DNA, two single-stranded DNA molecules were mixed at a 1:1 ratio in the annealing buffer (pH = 7.5), heated to 95°C for 5 min, and then slowly cooled to room temperature over 7 h.

The buffer used for imaging double-stranded DNA samples contained 50 mmol/L NaCl, 25 mmol/L

Tris-HCl, and pH at 7.5. The buffer used for imaging DNA Holliday junction samples contained 50 mmol/L NaCl, 50 mmol/L $MgCl_2$, 25 mmol/L Tris-HCl, and pH at 7.5. An oxygen-scavenging system was added to the buffer during single-molecule fluorescence imaging, which consisted of 0.8% D-glucose, 1 mg/mL glucose oxidase, 0.4 mg/mL catalase, and 1 mmol/L Trolox.

2.3.3 Fluorescence Imaging

First, 1 mg/mL biotin-modified bovine serum albumin (biotin-BSA) was injected into the sample chamber and incubated for 5 min, followed by rinsing with PBS buffer to remove unbound biotin-BSA. Then, 10 μg/mL streptavidin (SA) solution was added and incubated for 5 min, followed by rinsing with PBS buffer to remove unbound SA. Next, 100 pmol/L biotin-labeled fluorescently labeled DNA was added to the sample chamber and incubated for 5 min, followed by rinsing with PBS buffer to remove unbound DNA. Finally, a mixture of buffer and an oxygen-scavenging system was injected into the sample chamber for single-molecule fluorescence imaging. The experimental setup consisted of a total internal reflection fluorescence microscope and an EMCCD-based dual-channel fluorescence resonance energy transfer imaging system[29−31]. The EMCCD exposure time was set to 50 ms. ImageJ and Matlab software were used to record and analyze the intensity data.

3 Results and Discussion

3.1 XPS Analysis of GO Thermal Reduction

The LB technique can conveniently and inexpensively transfer monolayer GO onto coverslips to prepare sample chambers for single-molecule SIFA experiments[11]. The experimental coverslips can withstand temperatures up to 600°C without breaking or deformation. Based on this, coverslips modified with monolayer GO were placed directly in a vacuum tube furnace for baking to obtain coverslips modified with rGO, which were then used for SIFA experiments.

rGO with different degrees of reduction may differ in physical, chemical, and optical properties[32]. By measuring the surface chemical composition and bonding state of GO and rGO using XPS, the reduction of GO after tube furnace baking can be analyzed[33]. The size of the monolayer GO sheet is mostly on the micrometer scale[11], which is smaller than the detection range of XPS. Direct measurement of rGO on the coverslip would be interfered with by the glass component of the coverslip. Therefore, GO dispersion was prepared into GO thin films, two pieces of coverslips were used to sandwich the GO thin film, and the modified GO and rGO slides were baked together. XPS analysis was performed on the thermally reduced rGO thin film.

As shown in Fig. 2(a), the C1s spectrum of GO shows three peaks, corresponding to C–C ($sp^2$ hybrid carbon), C–O (hydroxyl and epoxy), and C=O (carbonyl) [27]. After baking and reduction at 300°C and 400°C for 2 h, the peaks of C–O and C=O decreased significantly (Fig. 2(b), (c)), indicating that most of the oxygen-containing groups had been removed. The XPS survey spectrum also shows that GO was well reduced after baking, as shown in the right column of Fig. 2. The C/O ratios of the unreduced GO, 300°C-2h-rGO, and 400°C-2h-rGO were 1.14, 2.48, and 3.29, respectively. This result is consistent with previous studies, indicating that the reduction degree of thermally reduced rGO mainly depends on the reduction temperature and reduction time[34,35]. The adjustment of reduction temperature and reduction time is continuous, and the reduction degree of rGO can be continuously controlled by setting different reduction parameters.

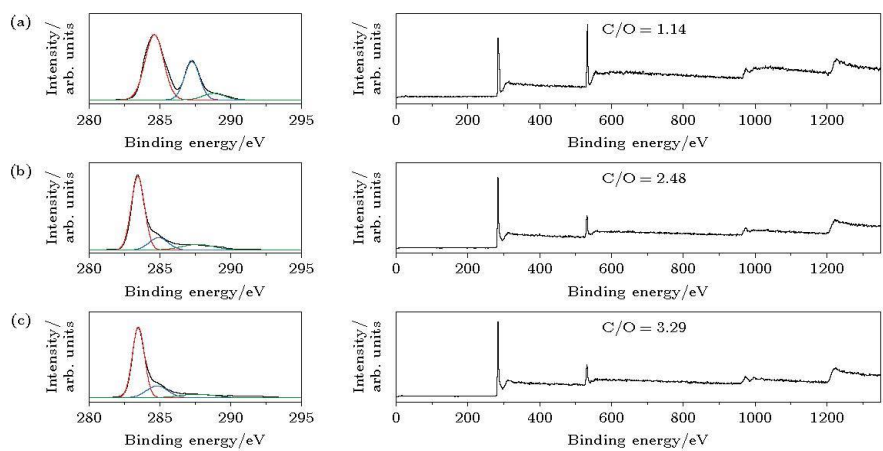

Fig. 2. XPS spectra of the original GO and rGO films. (a) c1s XPS spectra (left) and XPS survey spectra (right) for the original GO films; (b) c1s XPS spectra (left) and XPS survey spectra (right) for the 300℃-2h rGO films; (c) c1s XPS spectra (left) and XPS survey spectra (right) for the 400℃-2h rGO films

3.2 Determination of the Characteristic Quenching Distance $d_0$ of Thermally Reduced rGO

Before applying the rGO-based medium acceptor SIFA technique to study the vertical movement of biomolecules on the cell membrane surface, two issues need to be addressed: first, verify whether rGO can be used for the study of biomolecules; second, determine the $d_0$ of rGO. Double-stranded DNA has a stable structure in solution, and fluorescently labeled DNA can not only serve as a standard sample for single-molecule fluorescence imaging technology to verify the feasibility of the method but also be used to study proteins that interact with DNA. By connecting biotin-labeled double-stranded DNA to the rGO surface through biotin-BSA and SA, as shown in Fig. 3(a), the intensity of Cy3 can be measured and compared with the intensity measured on the glass surface to calculate the $d_0$ of rGO using Eq. (1). The left column of Fig. 3(b) shows the imaging of the sample chamber before the DNA was linked to the surface, and the right column shows the imaging after DNA linkage. After thermal reduction, rGO does not fluoresce spontaneously, and the fluorescence points cannot be distinguished from the rGO region or the glass region from the images. Therefore, it is necessary to analyze and statistics the intensity of all individual Cy3 molecules. The first row of Fig. 3(c)–(e) shows the intensity curves and statistical plots of individual Cy3 molecules labeled at the 1st bp, 9th bp, and 21st bp of the double-stranded DNA, respectively. The results show that the intensity of Cy3 is stable when labeled on the double-stranded DNA and is independent of the labeling site. The intensity of Cy3 follows a Gaussian distribution, and the peak is $I_0$. The second and third rows of Fig. 3(c) show the intensity curves and intensity distributions of Cy3 labeled at the 1st bp of DNA imaged in the 300°C -2h-rGO and 400°C-2h-rGO sample chambers, respectively. In the rGO sample chamber, Cy3 exhibits two stable intensities: a higher intensity that is the same as that imaged on the glass, indicating that the DNA is connected to the glass; a lower intensity due to the fluorescence quenching effect of rGO on Cy3, indicating that the DNA is connected to the rGO. The intensity curves of the lower intensity are stable and do not fluctuate, indicating that the height of Cy3 from the rGO surface remains constant.

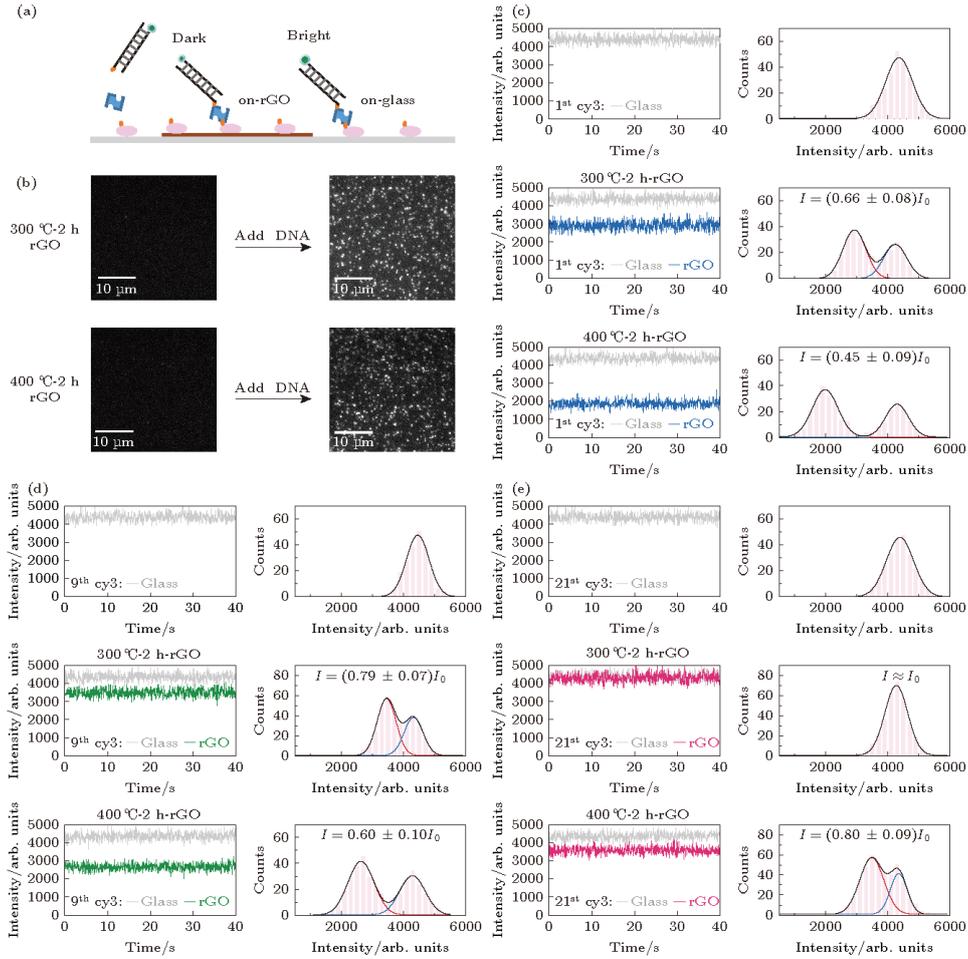

Fig. 3. Determination of $d_0$ of rGO by fluorescence labeled DNA. (a) Schematic representation of DNA imaging; (b) DNA imaging on 300 ℃-2h rGO (upper) and 400 ℃-2h rGO (lower); (c-e) Intensities of cy3 labeled at 1bp (c)、9 bp (d)、21bp (e) of DNA on glass and rGO.

There is a difference in the lower intensity of Cy3 in the 300°C-2h-rGO and 400°C-2h-rGO sample chambers, which are $0.66I_0$ and $0.45I_0$, respectively, indicating that the fluorescence quenching effect of 400°C-2h-rGO on fluorescence is stronger than that of 300°C-2h-rGO. The intensity of Cy3 labeled at the 9th bp of DNA was imaged and analyzed for individual Cy3 intensity in the 300°C-2h-rGO sample chamber, as shown in Fig. 3(d). The intensity of Cy3 labeled at the 9th bp is stronger than that labeled at the 1st bp, which is due to the fact that the distance of Cy3 from the rGO surface is further away than that at the 1st bp. When Cy3 labeled at the 21st bp of DNA was imaged in the 300°C-2h-rGO sample chamber, there was only one intensity for each single Cy3 molecule, and it was close to the intensity measured in the glass sample chamber, indicating that the fluorescence quenching effect of rGO on Cy3 was not significant at this time. The statistical results show that the intensity has only one peak, but the peak width is wider than the $I_0$ distribution measured on the glass. When the DNA with a Cy3 label at the 21st bp was imaged in the 400°C-2h-rGO sample chamber, it still showed two intensity distributions, indicating that 400°C-2h-rGO has a stronger fluorescence quenching effect.

The residence length of double-stranded DNA in solution is 50 nm[36], and the distance between adjacent base pairs is 0.34 nm[37]. Therefore, the 21 bp double-stranded DNA used in the experiment can be approximately regarded as a rod-like rigid body with a length of 7.1 nm. When calculating the height of Cy3 from the rGO surface, only the angle between the double-stranded DNA and the rGO surface in the vertical direction needs to be considered. Studies have shown that the angle between short double-stranded DNA fixed on the surface and the surface normal is constant and is about 60°[38,39]. Combined with previous studies that indicate that the size of BSA is about 3 nm[40], and the size of SA is about 4.2 nm[41], the normal distances of Cy3 labeled at the 1st bp, 9th bp, and 21st bp from the rGO surface can be calculated. Table 1 summarizes the $d_0$ calculated using the intensity and distance information of different fluorescently labeled DNAs imaged on glass and rGO. In the same rGO sample chamber, $d_0$ was calculated by imaging and calculating the $d_0$ of DNA with different fluorescent labeling positions. The results were close, further confirming the feasibility of the rGO-SIFA method. Imaging DNA with different fluorescent labeling positions on the same rGO sample chamber and calculating $d_0$ belongs to an independent experiment. The results of independent experiments can be combined to obtain more accurate results. By calculating the weighted average, $d_0$ of 300°C-2h-rGO is (6.3 ± 0.5) nm, and $d_0$ of 400°C-2h-rGO is (7.9 ± 0.5) nm. Substituting the calculated $d_0$ of 300°C-2h-rGO and the height of Cy3 labeled at the 21st bp into Eq. (1), the theoretical intensity of Cy3 can be calculated to be approximately $0.9I_0$. Due to the differences between individual molecules and the error of instrumental measurement, the intensity of Cy3 follows a Gaussian distribution with a certain peak width. Even if the intensity distribution has two peaks of $0.9I_0$ and $I_0$, it cannot be distinguished, which is consistent with the experimental results. The difference in $d_0$ between 300°C-2h-rGO and 400°C-2h-rGO is a predictable result. By reducing the reduction temperature, $d_0$ can gradually transition from 6.3 nm to 4 nm. The highest reduction temperature used in this study is 400°C, which is still far from the high-temperature limit of 600°C for the glass cover glass used in the experiment. On the other hand, the glass cover glass can also be replaced with quartz cover glass to achieve higher temperature thermal reduction, further increasing $d_0$.

Table 1. Determination of $d_0$ of rGO by fluorescence labeled DNA.

|  | 1 bp (7.5 nm) | 9bp (8.9nm) | 21bp (10.9 nm) |
| --- | --- | --- | --- |
| 300°C-2h-rGO | $I = 0.66 \pm 0.08\ I_0$ | $I = 0.79 \pm 0.07\ I_0$ | $I \approx I_0$ |
|  | $d_0 = 6.4 \pm 0.7$ nm | $d_0 = 6.2 \pm 0.6$ nm |  |
| 400°C-2h-rGO | $I = 0.45 \pm 0.09\ I_0$ | $I = 0.60 \pm 0.10\ I_0$ | $I = 0.80 \pm 0.09\ I_0$ |
|  | $d_0 = 7.9 \pm 0.7$ nm | $d_0 = 8.1 \pm 0.8$ nm | $d_0 = 7.8 \pm 0.9$ nm |

3.3 Measuring DNA Holliday Junction Conformational Changes with rGO-SIFA

When different $d_0$ acceptor materials are used in single-molecule SIFA technology, it is possible to detect the dynamic processes of biomolecules at different heights from the cell membrane surface. Using fluorescently labeled double-stranded DNA to measure the $d_0$ of rGO, the DNA attached to the rGO shows only one intensity of light, meaning that the target biomolecule is in only one state. To better verify the feasibility of the rGO-SIFA method, it is necessary to select a biomolecule whose fluorescence-labeled site changes in height from the rGO for study. This biomolecule should have a stable structure in solution and have been well-researched. Holliday junction is an important intermediate in DNA replication and homologous recombination, consisting of four DNA strands[42]. In the absence of divalent metal ions in solution, Holliday junction presents a static cross-shaped structure, while in the presence of divalent metal ions, Holliday junction will stack into an X-shaped structure[43]. The X-shaped structure of Holliday junction has two different conformations, and these two conformations are constantly transforming into each other, with the transformation rate regulated by the concentration of divalent metal ions[44,45]. The conformational transition of Holliday junction is a necessary condition for homologous recombination, and the nucleotide sequence in the central region of Holliday junction will affect the transition rate and conformational persistence time of the two conformations[45,46]. Due to the simple structure and ease of preparation of Holliday junction, and the stable observation of the mutual transition of the two conformations in solution, fluorescently labeled Holliday junction is often used to verify advanced single-molecule fluorescence imaging techniques[47,48]. Construct a Holliday junction as shown in Fig. 3, Table 2 shows the nucleotide sequences of the four short DNA strands that make up the Holliday junction, as well as the Cy3 labeling position. The Holliday junction with this sequence can be connected to the cover glass surface and has a stable conformational transition in the presence of divalent metal ions[46,48].

Table 2. Nucleotide sequences of DNA Holliday junction.

| Strand | Nucleotide sequence |
|--------|---------------------|
| X | CCC AGT TGA GAG CTT GAT AGG G |
| B | CCC TAT CAA GCC GCT GTT ACG G |
| R | CCC ACC GCT CTT CTC AAC TGG G |
| H | biotin-CCG TAA CAG CGA GAG CGG TGG G(cy3) |

Cy3 is labeled at the 3' end of the H strand, and the Holliday junction is fixed to the surface by the biotin labeled at the 5' end of the H strand. Under the conditions of 50 mmol/L $Mg^{2+}$ and 50 mmol/L $Na^+$, Holliday junction undergoes continuous conformational changes. In state 1, Cy3 is close to the surface of rGO/GO or glass (approximately 7.5 nm), and in state 2, the distance is far.

Fig. 4(a) shows the results of observing Holliday junction in the glass sample chamber, with the left column displaying the intensity of a single Cy3 molecule. The intensity is stable and does not fluctuate, and the statistical results show that the intensity of Cy3 is a Gaussian distribution, with the central value being $I_0$. Fig. 4(b) shows the observation of Holliday junction in the GO sample chamber. No change in the intensity of Cy3 is observed. The closest distance of Cy3 to the surface is approximately 7.5 nm, and the $d_0$ of GO=4 nm. By substituting (1) formula, the theoretical intensity of Cy3 when it is close to GO can be calculated as $0.94I_0$. The distance of 7.5 nm has exceeded the sensitive range of GO. At this time, as the distance increases, the intensity gradually approaches $I_0$ from $0.94I_0$, and this slight change in intensity is already below the sensitivity of the instrument. The statistical results of Cy3 intensity on GO are Gaussian distribution, but the full width at half maximum is larger than that on glass. There are two reasons for this phenomenon: one is that the theoretical calculation shows that Cy3 on GO has two intensities, and both of them are close to $I_0$, making it difficult to distinguish the two intensities in the experiment; the other is that GO itself emits weak fluorescence, and the signal-to-noise ratio of single molecules imaged on GO is poorer than that on glass, so the peak width of the intensity distribution will be larger than that on glass. Fig. 4(c) shows the observation of Holliday junction in the 400°C-2 h-rGO sample chamber.

The intensity of Cy3 jumps between two high and low values. When observing Cy3 labeled double-stranded DNA on 400°C-2 h-rGO, the intensity of Cy3 is stable and does not fluctuate, indicating that 400°C-2 h-rGO only has a quenching effect on the intensity of Cy3 but does not affect the stability of the intensity of Cy3. Therefore, the change in the intensity of Cy3 labeled by Holliday junction is due to the change in the height of Cy3 from 400°C-2 h-rGO. By time-weighted statistical analysis of the intensity of Cy3, the results show that the intensity of Cy3 presents a distribution with two peaks, with peak values of $0.42I_0$ and $0.83I_0$, respectively. By substituting (1) formula and $d_0 = (7.9 \pm 0.5)$ nm of 400°C-2 h-rGO, the normal distances of Cy3 from rGO are calculated to be 7.3 nm and 11.7 nm, which are close to the theoretical distances of 7.5 nm and 11.1 nm of Cy3. According to the intensity distribution diagram of Cy3 in Fig. 4(c), the peak height ratio of high and low intensities can be calculated to be 1:2, indicating that the duration of state 1 conformation of Holliday junction is twice that of state 2, which is consistent with the previous research results on Holliday junction with the same DNA sequence[46]. Based on the above two points, it is confirmed that the conformational transition of Holliday junction is observed using 400°C-2 h-rGO, and this conformational transition cannot be observed by GO. The reason for the difference in results is that 400°C-2 h-rGO has a larger $d_0$ than GO, allowing observation of a wider range. The thermally reduced rGO can adjust $d_0$ by controlling the reduction temperature, and the appropriate reduction temperature should be set flexibly according to the scale of the target macromolecule and the fluorescence labeling site to select a suitable $d_0$ for observation.

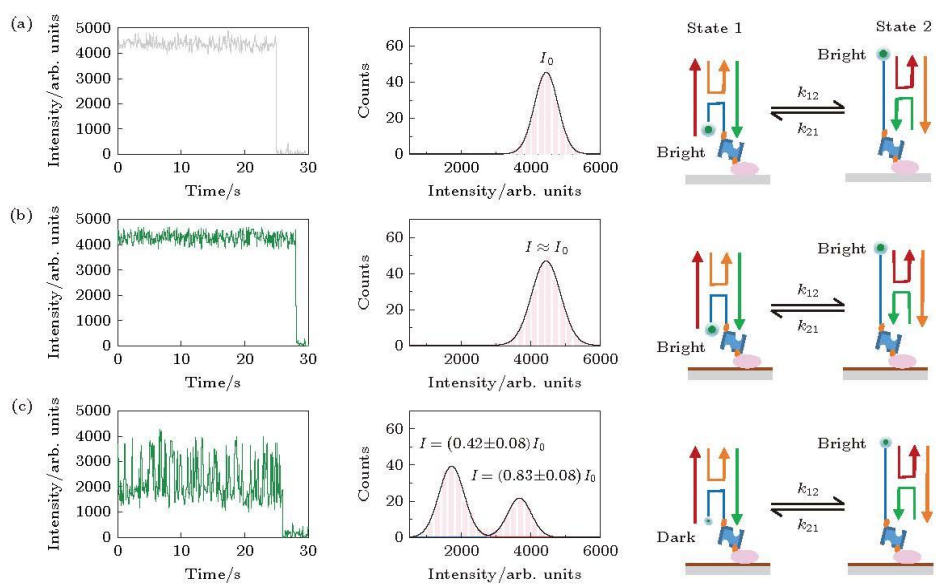

Fig. 4. Observing the conformational transition of Holliday junction by SIFA. (a-c) Observing the Cy3 labeled Holliday junction on glass (a)、GO (b)、400℃-2h rGO (c), left columns show intensity-time curves of a single Cy3, middle columns show distribution of intensities of Cy3, right columns show schematic representation of the change of Cy3 light intensity caused by the conformational transition of Holiday junction.

## 4 Conclusion

This study advances our group's research on the Surface-Induced Fluorescence Attenuation (SIFA) technique. By baking cover glass modified with GO in a vacuum tube furnace at controlled temperatures, we can achieve different degrees of reduced graphene oxide (rGO). This process allows us to adjust the characteristic quenching distance ($d_0$), broadening the applicability of this method for single-molecule studies in two-dimensional systems. Specifically, the $d_0$ for 300°C-2h-rGO was measured at 6.3 ± 0.5 nm, and for 400°C-2h-rGO at 7.9 ± 0.5 nm using fluorescently labeled DNA. We anticipate that adjusting the reduction temperature between room temperature and 400°C will enable continuous fine-tuning of $d_0$ from 4 nm to 7.9 nm. Observations of Cy3-labeled Holliday junctions on both GO and 400°C-2h-rGO revealed that while GO could not detect any intensity change of Cy3, the 400°C-2h-rGO displayed fluctuating high and low intensity values. These intensity variations, calculated against the rGO surface distance, closely matched theoretical expectations, underscoring the enhanced measurement accuracy and spatial resolution of SIFA when based on rGO. Importantly, since rGO itself does not emit fluorescence, it does not interfere with the signal-to-noise ratio during single-molecule fluorescence imaging. By integrating SIFA with Fluorescence Resonance Energy Transfer (FRET) techniques, we can now observe the three-dimensional movements and conformational changes of biomolecules in real time. Looking forward, the rGO-SIFA technique holds significant potential for broader applications in studying the function and structure of membrane proteins.